# Investigation of the electronic structure of biphenylene ribbons of the armchair type

Gennadiy Ivanovich Mironov*[a]

[a] Prof. G.I. Mironov
Department of Physics and Materials Science
Mari State University
Kremlevskaya Street 44, Yoshkar-Ola (Russia)
E-mail: mirgi@marsu.ru

**Abstract:** The electronic structure of chair-type biphenylene ribbons is studied within the framework of the Hubbard Hamiltonian in the approximation of static fluctuations. An expression is obtained for the Fourier transform of the Green's functions, whose poles determine the spectrum of elementary excitations. The energy spectrum of biphenylene ribbons indicates that nanoribbons with a width of 6, 9, 12, 15, 18 carbon atoms are in the semiconductor state. Densities of state of electrons of biphenylene ribbons of various widths are presented in comparison with the results of experimental measurements of differential conductance of biphenylene ribbons. It is shown that the width of the band gap decreases exponentially with an increase in the width of the biphenylene ribbon. Ribbons with a width of 21 carbon atoms or more are in the metallic state.

## 1. Introduction

The last thirty-five years testify to a giant breakthrough in the synthesis and study of new allotropic states of carbon, their application in various branches of science and technology. The desire to understand the mysterious nature of the optical absorption spectrum in the atmosphere of stars led to the discovery of fullerene [1], the process of synthesis of carbon fullerenes initiated the discovery of carbon nanotubes [2], the search for the possibility of separating a layer of carbon atoms one atom thick from graphite led to the discovery of graphene as a qualitatively new allotropic state of carbon [3, 4]. Recently, an article [5] was published, the authors of which obtained a planar carbon structure of the graphene type, consisting not only of hexagons, but also of regularly arranged eight-membered and four-membered rings. They synthesized a biphenylene network, which demonstrates the transition from a semiconductor state to a metallic state as the size of the network increases.

Soon after the discoveries of fullerene, nanotubes, and graphene, papers [6–10] were published, which made it possible to mainly explain the physicochemical properties of these allotropic states of carbon. Questions remained however. For example, in the case of $C_{60}$ fullerene, the question remained open about the optical spectrum of fullerene [11-13]. The experimentally observed optical absorption spectrum of $C_{60}$ fullerene was explained under the assumption that the system of $\pi$ electrons in fullerene is a strongly correlated system; see, for example, [14]. In the case of nanotubes, the question arose of metallic and semiconductor conductivity. In works [7-9], it was theoretically shown that the metallic type of the structure of the energy bands of nanotubes is obtained if the difference of chiral indices (n-m) is a multiple of three, the rest of the nanotubes must have the properties of semiconductors. Soon this quantum-chemical prediction was confirmed [15-18], but the experimental results [15-18] did not have a decisive effect, since these studies did not achieve complete control over the structure of the nanotubes under study. In [19], at a temperature of 5 K, for individual ultrapure nanotubes (9, 0), differential conductivities near the Fermi energy were determined by scanning tunneling microscopy, which are equal to the density of electronic states to within a constant factor. It turned out that there was a narrow energy gap in the Fermi energy region, which meant that the "metal" carbon nanotube was not actually metallic. Experiments [20] confirmed the presence of narrow energy bands. Therefore, zigzag-type nanotubes, regardless of the difference in chiral indices, are narrow-gap semiconductors and not metals. In [21], taking into account the tight coupling model [22], the electron-electron correlation of $\pi$-electrons [23, 24], and the finite curvature of the nanotube surface [19, 22] the electronic structure of the nanotube (9, 0) was calculated. The resulting width of the band gap coincided with the experimental value [19]. In connection with the impressive results of [5], it is necessary to understand how a carbon nanosystem, which includes four-membered, six-membered, and eight-membered rings, can transform into a metallic state as it grows, since in the case of graphene, an increase in size only leads to the disappearance of the energy gap between the conduction band and the valence band.

In each of the considered cases of new allotropic states of carbon (fullerene, nanotubes, graphene, biphenylene), each carbon atom is surrounded by three nearest neighboring carbon atoms. S-p hybridization in the case of these states leads to the fact that the three σ-electrons of each carbon atom form three covalent bonds with three neighboring atoms. The energy level of the fourth π-electron is located above the energy level of the σ-electrons, and the wave function of the π-electron overlaps with the wave functions of all three neighboring π-electrons. For this reason, the π-electron can go to any of the three neighboring atom. The π-electrons are responsible for the electronic and transport properties of the considered carbon nanosystems, at least they make the main contribution to these properties. If a π-electron has passed to a neighboring atom in accordance with the Pauli principle, then two π-electrons located in the same orbital will experience a strong Coulomb repulsion from each other, the energy of their Coulomb repulsion will be much greater than the π-electron transfer integral. Therefore, in the carbon nanosystems under study, a system of strongly correlated π-electrons will appear, which is responsible for the main properties of carbon nanosystems. A strongly correlated system has a number of features [25–27]. The most important feature is that in the case when there is on average one π-electron for each site of the carbon nanosystem, the band of valence electrons is completely filled. Accounting for this feature in the above works of the author made it possible to explain both the optical absorption spectra of fullerenes and the presence of narrow bands of forbidden energies in "metallic" carbon nanotubes.

## 2. Theoretical model

When moving from real biphenylene networks to building a model, taking into account the above, it should be taken into account that π-electrons can move from one carbon atom (biphenylene site) to another neighboring site, that two π-electrons located at the same biphenylene side will experience a strong Coulomb repulsion. Since π-electrons are responsible for the main properties in carbon nanosystems, the influence of σ-electrons on the properties of carbon nanosystems can be neglected, modeling can be performed within the framework of the single-band Hubbard Hamiltonian [28–29].

The Hamiltonian of the Hubbard model, which consists of N atoms of a biphenylene ribbon, can be written as:

$$\widehat{H} = \varepsilon \sum_f^N (\hat{n}_{f\uparrow} + \hat{n}_{f\downarrow}) + \sum_{f \neq f', \sigma} B_{f,f'} \left( a_{f\sigma}^+ a_{f'\sigma} + a_{f'\sigma}^+ a_{f\sigma} \right) + U \sum_f^N \hat{n}_{f\uparrow} \hat{n}_{f\downarrow}, \qquad (1)$$

where $\varepsilon$, $B_{f,f'}$, $U$ are the self-energy, the transfer integral and the Coulomb potential, respectively. $\hat{n}_{f\uparrow}$ is the operator of the number of π-electrons at the site f of biphenylene with spin projection ↑ (σ=↑,↓), $a_{f\sigma}^+$ is the operator of π-electron creation, $a_{f\sigma}$ is the annihilation operator.

In [5], biphenylene chair-type ribbons 6 atoms wide (6-BPR), as well as 9-BPR, 12-BPR, 15-BPR, 18-BPR, and 21-BPR were obtained and studied. If the first five biphenylene ribbons were in the semiconductor state, with the width of the band gap successively decreasing as the width of the ribbon increased, then 21-BPR was in the metallic state. The purpose of our work is to find a possible explanation for the semiconductor-metal transition in [5].

If we build a model of a biphenylene ribbon 6 atoms wide, consisting of 108 carbon atoms, and number all the nodes of the ribbon, then for the electron production operators in the Heisenberg representation, we can obtain the following system of 108 equations of motion (see fig. 1, where the 6-BPR biphenylene tape is shown with some site numbering):

$$\frac{d}{d\tau} a_{1\uparrow}^+ = \varepsilon a_{1\uparrow}^+ + B(a_{2\uparrow}^+ + b a_{27\uparrow}^+ + a_{45\uparrow}^+) + U \hat{n}_{1\downarrow} a_{1\uparrow}^+,$$

$$\frac{d}{d\tau} a_{2\uparrow}^+ = \varepsilon a_{2\uparrow}^+ + B(a_{1\uparrow}^+ + b a_{28\uparrow}^+ + a_{46\uparrow}^+) + U\hat{n}_{2\downarrow} a_{2\uparrow}^+, \qquad (2)$$

$$\ldots\ldots\ldots\ldots\ldots\ldots\ldots\ldots\ldots\ldots\ldots\ldots\ldots\ldots\ldots\ldots\ldots\ldots\ldots$$

$$\frac{d}{d\tau} a_{108\uparrow}^+ = \varepsilon a_{108\uparrow}^+ + B(a_{72\uparrow}^+ + a_{107\uparrow}^+) + U\hat{n}_{108\downarrow} a_{108\uparrow}^+.$$

In the system of equations (2) $\tau = it$ is imaginary time. Let us pay attention to the factor b in equations (2). From the analysis of Fig. 2 it follows that the four-membered ring is not a square. The distance between sites 1 and 27 is less than the distance between atoms 1 and 2. And this means that the wave functions of atoms 1 and 27, 2 and 28 overlap much more than, for example, atoms 1 and 2. The transfer integral depends on the degree of overlapping of the wave functions, and a relatively insignificant change in the degree of overlapping of the wave functions of interacting atoms leads to a very significant difference in the transfer integrals (see, for example, [19, 21, 22]). The parameter b just takes into account the change in the value of the transfer integral.

Let us solve the system of equations of motion (2) in the approximation of static fluctuations [30–32]. The method for solving systems of equations of motion of type (2) is given in previously published works [30-32], for this reason, without giving details, we present the final result.

Atom number one is chosen in the center of the nanoribbon, this node is common for four-, six-, eight-membered rings. Having solved the system of equations (2), we calculate the Fourier transform of the anticommutator Green's function, the general form of the solution is represented as follows (in the case of the atom number one):

$$\langle\{a_{1\uparrow}^+|a_{1\uparrow}\}\rangle_E = \frac{i}{2\pi} \sum_{j=1}^{107} \left\{ \frac{W_j}{E - V_j - \varepsilon} + \frac{W_j}{E - V_j - \varepsilon - U} \right\}. \qquad (3)$$

The poles of the Fourier transform of the Green's function (3) determine the energy spectrum. If we take a specific value of j, setting the denominator of the fraction equal to zero, we determine the energy level of the π-electron, the numerator of this fraction will determine the probability of finding the π-electron at this energy level. We will not write out the numerical value of each energy level and the probability of finding an electron at this energy level in the form of a table, since the table will have a very cumbersome form. All 107 energy levels are given in the form of an energy spectrum, and the first fraction in (3) with the value of $\varepsilon$ in the denominator will determine the valence band, the second fraction with the value of $\varepsilon + U$ will determine the conduction band. The numerator of the fraction in the Green's function in formula (3) will affect the density of the electronic state.

## 3. Results and Discussion

In Figure 3, the energy spectrum of 6-BPR of 108 carbon atoms is shown for the following parameter values: $U = 7\ eV, B = -1\ eV, b = 1.74, \varepsilon = -U/2$. These parameters were chosen using optimization methods based on the values known in the literature; they made it possible to explain the optical absorption spectra [14] and the values of the energy gaps in carbon nanosystems [21]. On fig. 3, a set of 107 energy levels above 0 eV forms the conduction band (upper Hubbard subband), the lower levels form the valence band (lower Hubbard subband). The width of the band gap is equal to $\Delta = 0.47$ eV. For comparison, we note that the energy gap in the case of fullerene $C_{60}$ $\Delta = 1.557$ eV [14].

Figure 4 shows the electron density of state in the case of 6-BPR of 108 carbon atoms at system parameters $U = 7\ eV, B = -1\ eV, C = 0.02\ eV, \varepsilon = -U/2$, where C is the half-width used for modeling the delta function. If we take the two peaks closest to 0 eV, which correspond to Van Hove singularities, we get $\Delta = 0.47\ eV$. Thus, 6-BPR of 108 carbon atoms is in the semiconductor state. It follows from the analysis of Figure 3A in [5] that the differential conductance measurement carried out in the position marked in Fig. 3B shows that the band gap is $E_g = 2.35\ eV$. An analysis of the energy gap value depending on the number of atoms shows that in carbon nanosystems, as the number of carbon atoms increases, the energy gap width decreases exponentially. We noted above that in the case of 60 carbon atoms the gap width is of the order of one and a half electron volts. As the number of carbon atoms in the nanosystem increases, the width of the energy gap will gradually decrease. Note that if we measure the differential conductance, which characterizes the density of the electronic state, in a different position, for example, associated with the 45th atom (see Fig. 1), then the electron state density will have the form shown in Figure 5. Based on how the energy gap was determined for 15-BPR in Fig. 3A, we could assume that the band gap is equal to $2.66\ eV$, which is close in order of magnitude to the value of $E_g = 2.35\ eV$.

The bandgap evolution of the BPRs is shown in Fig. 6. Consider a 9-BPR biphenylene ribbon containing 162 carbon atoms. The energy spectrum of 9-BPR is shown in Figure 6A on the left, the electron state density is shown in this figure on the right. The figure shows only part of the energy spectrum near the Fermi level (0 K), since we need to see how the width of the forbidden energy band decreases as the width of the biphenylene ribbon increases. In the case of 9-BPR, the band gap is $\Delta = 0.19\ eV$. In [5], the 9-BPR energy gap has a width of $E_g = 1.44\ eV$. The density of state shows that 9-BPR is a good semiconductor.

Consider next the 12-BPR biphenylene ribbon containing 216 carbon atoms (Fig. 6b). The width of the energy gap is equal to $\Delta = 0.09\ eV$, in [5] the energy gap has a width of $E_g = 0.84\ eV$. In Figure 6C, the energy spectrum and electronic state density are shown when 15-BPR consists of 270 carbon atoms. The width of the energy gap is equal to $\Delta = 0.04\ eV$, in [5] the energy gap has a width of $E_g = 0.70\ eV$. On fig. 6D we have the case of a biphenylene ribbon 18 atoms width with a total of 324 carbon atoms (18-BPR). The width of the energy gap is equal to $\Delta = 0.014\ eV$, in [5] the energy gap has a width of $E_g = 0.36\ eV$. It should be noted that qualitatively the behavior of the curves for the differential conductance in [5] and the electron state density in Fig. 6 coincide, especially in the case of 18-BPR. Before the transition to the metallic state, a Van Hove singularity begins to form near the Fermi level.

In Figure 6E, the energy spectrum and electron state density are shown in the case when 21-BPR consists of 378 carbon atoms. The conduction band and the valence electron band are overlapped, and 21-BPR is in the metallic state. The absence of an energy gap in the spectrum and the finite value of the density of state at the Fermi level indicate metallicity. The Van Hove singularity is at the Fermi level. Thus, as the width of the biphenylene ribbon increases, the semiconductor-metal transition occurs, the energy gap is completely closed between 18-BPR and 21-BPR.

On Fig. 7A according to the data given in [5], the dependence of the width of the energy gap $E_g$ on the width of the biphenylene tape is shown, the solid graph represents an exponential curve constructed by the optimization method (approximation). On Fig. 7B a similar dependence of the energy gap width (this value is denoted as $\Delta$ in the work) from the width of the biphenylene ribbon is shown. The solid curve is an exponential interpolation curve. The width of the energy band decreases exponentially as the width of the biphenylene ribbon increases.

## 4. Conclusion

Thus, it is possible to describe the semiconductor–metal transition for $\pi$-electrons in a biphenylene network within the framework of the single-band model. Unlike graphene, the biphenylene network contains adjacent four- and eight-membered rings. This leads to the fact that the distance between some sites of the network becomes smaller, which leads to a greater degree of overlap of the wave functions of these atoms. As a consequence, the electron transfer integral between these atoms becomes larger, resulting in a semiconductor-metal transition as the biphenylene network grows. The presence of biphenylene networks in both the semiconductor and metallic states opens up great prospects for the practical application of this new state of carbon nanosystems. As noted in [5], the metallicity of the biphenylene network makes it a good candidate for the role of conductors in future carbon-based electronic circuits, especially in those areas where intensive cooling of electronic devices is required. The metallicity is due to the presence of a sufficient number of four-membered rings in biphenylene networks. Therefore, close attention should be paid to studying the effect of four-membered rings on the physicochemical properties of biphenylene networks.

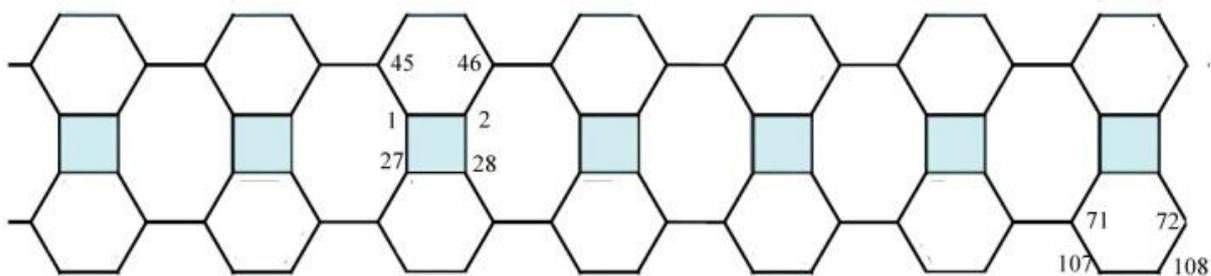

**Figure 1.** Biphenylene ribbon 6-BPR with numbering of the location of some atoms

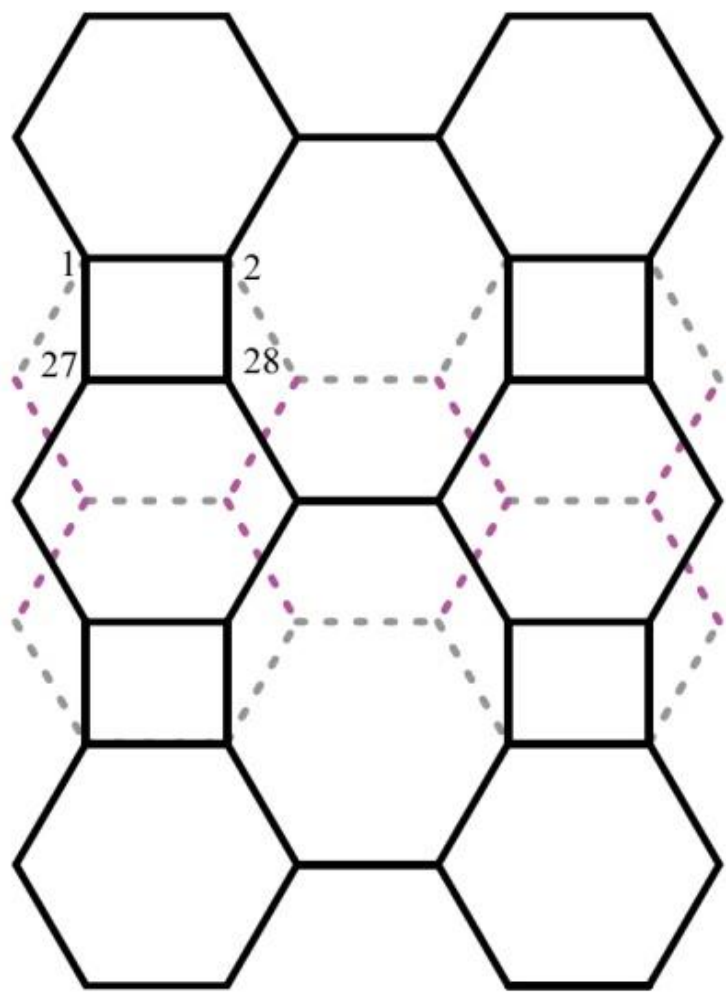

**Figure 2.** Fragment of a biphenylene network

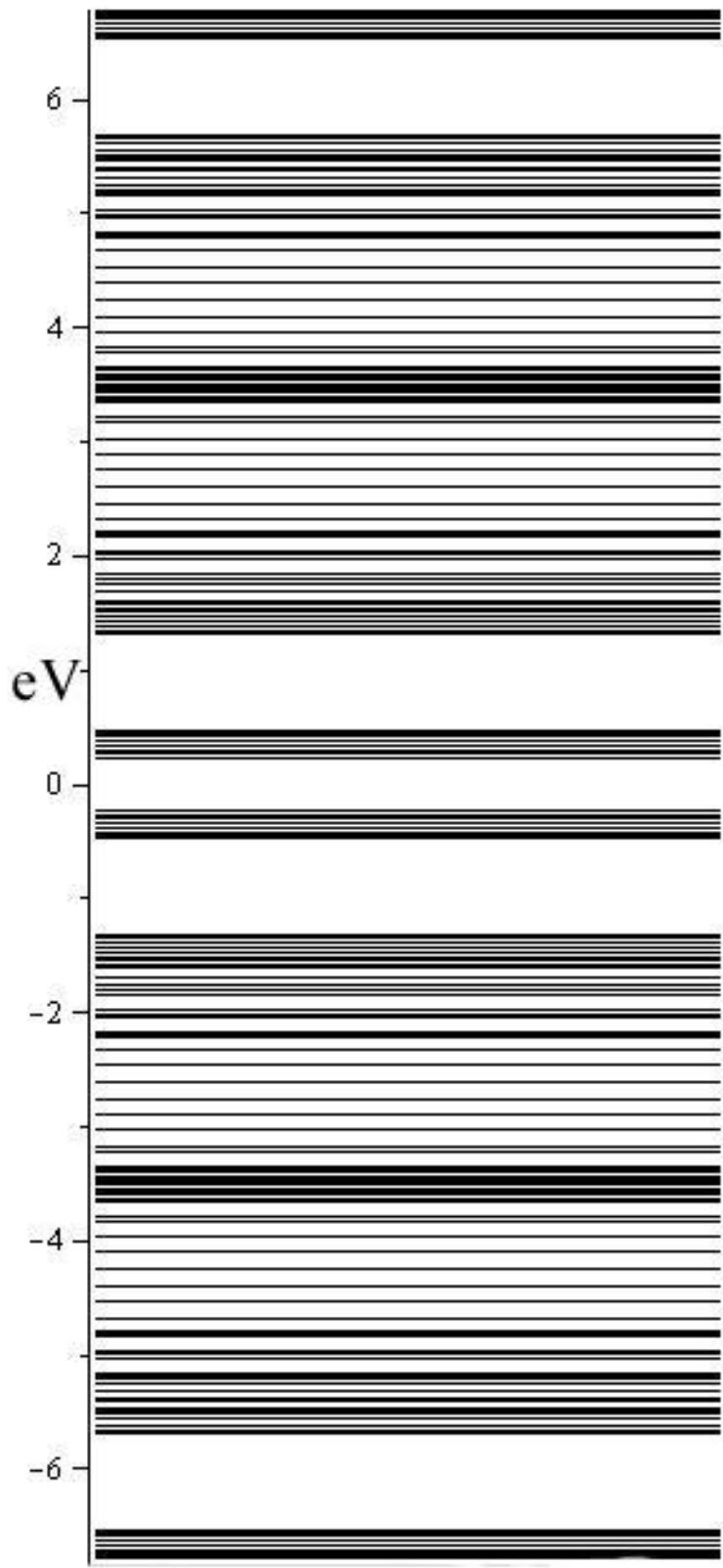

**Figure 3.** Energy spectrum of 6-BPR at the parameter values: $U = 7\ eV, B = 1\ eV, b = 1.74, \varepsilon = -U/2$.

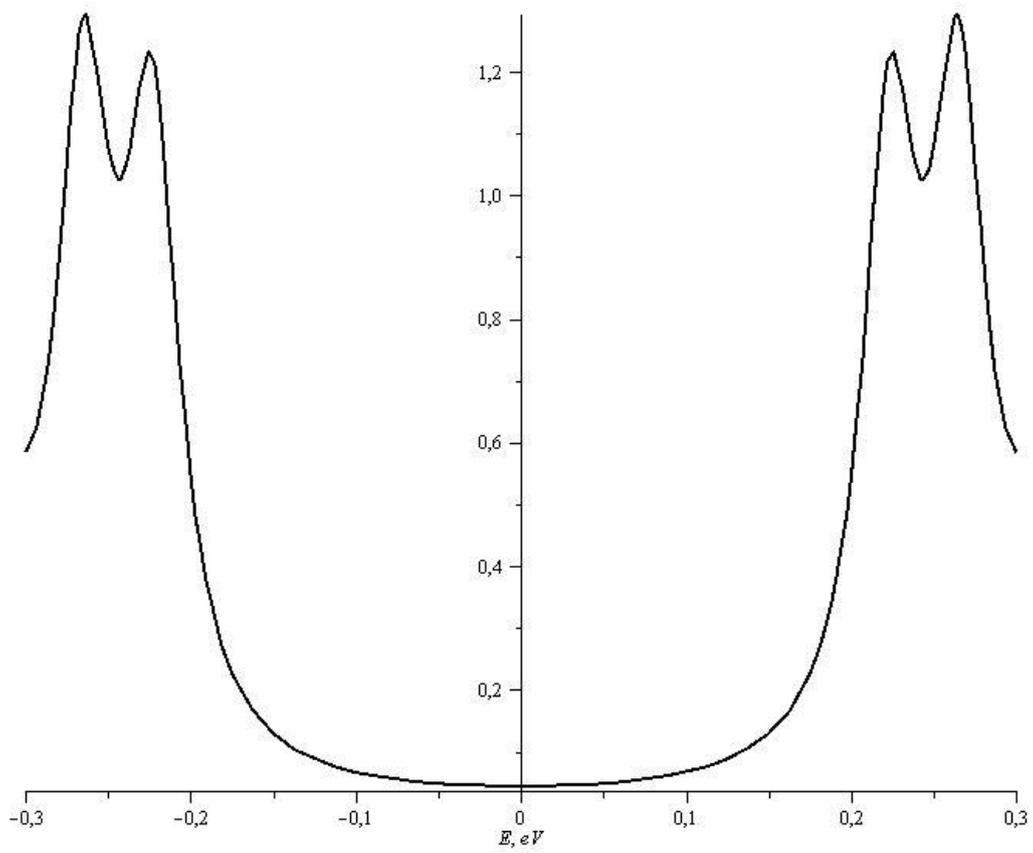

**Figure 4.** Density of state of electrons in arbitrary units for the values of the parameters: $U = 7\ eV, B = 1\ eV, C = 0.02\ eV, \varepsilon = -U/2$

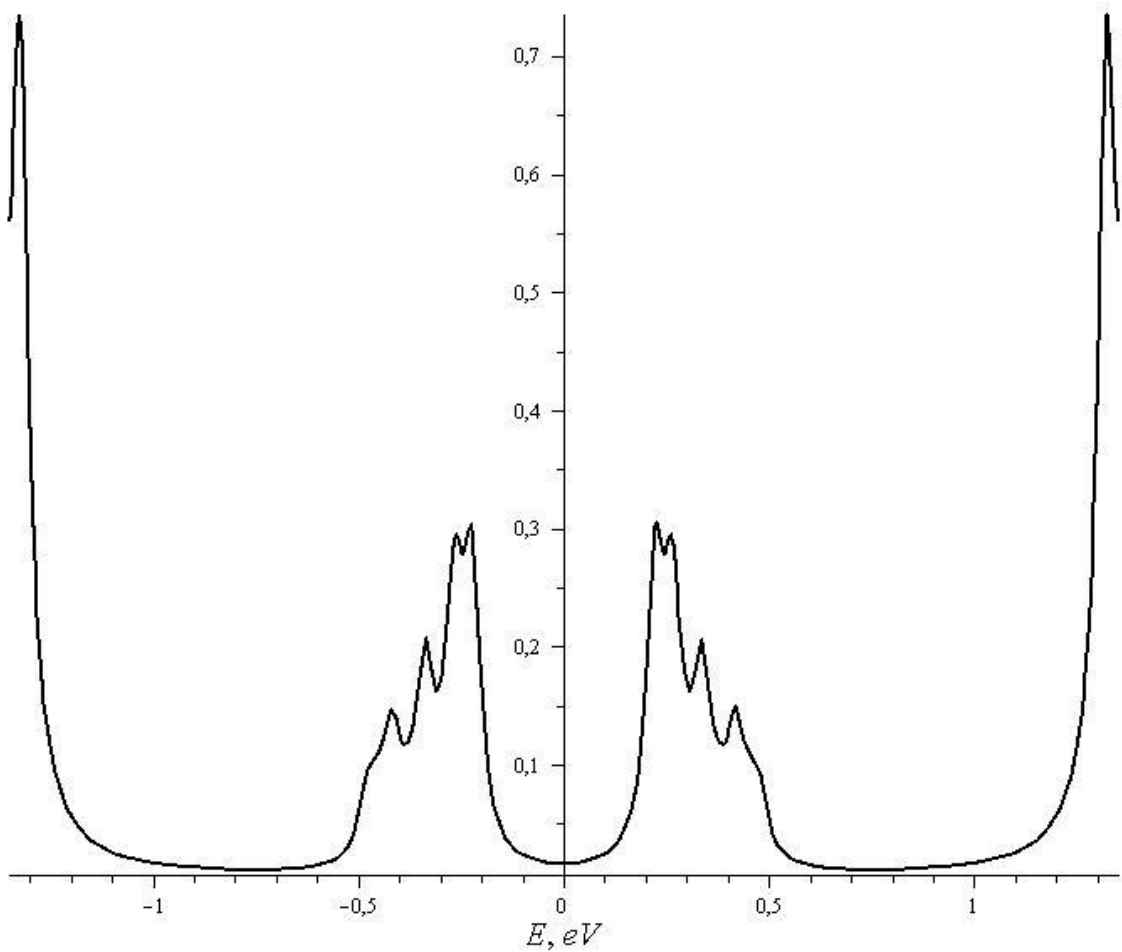

**Figure 5.** Density of state of electrons in arbitrary units at parameter values: $U = 7\ eV, B = 1\ eV, C = 0.02\ eV, \varepsilon = -U/2$ in the case of atom number 45

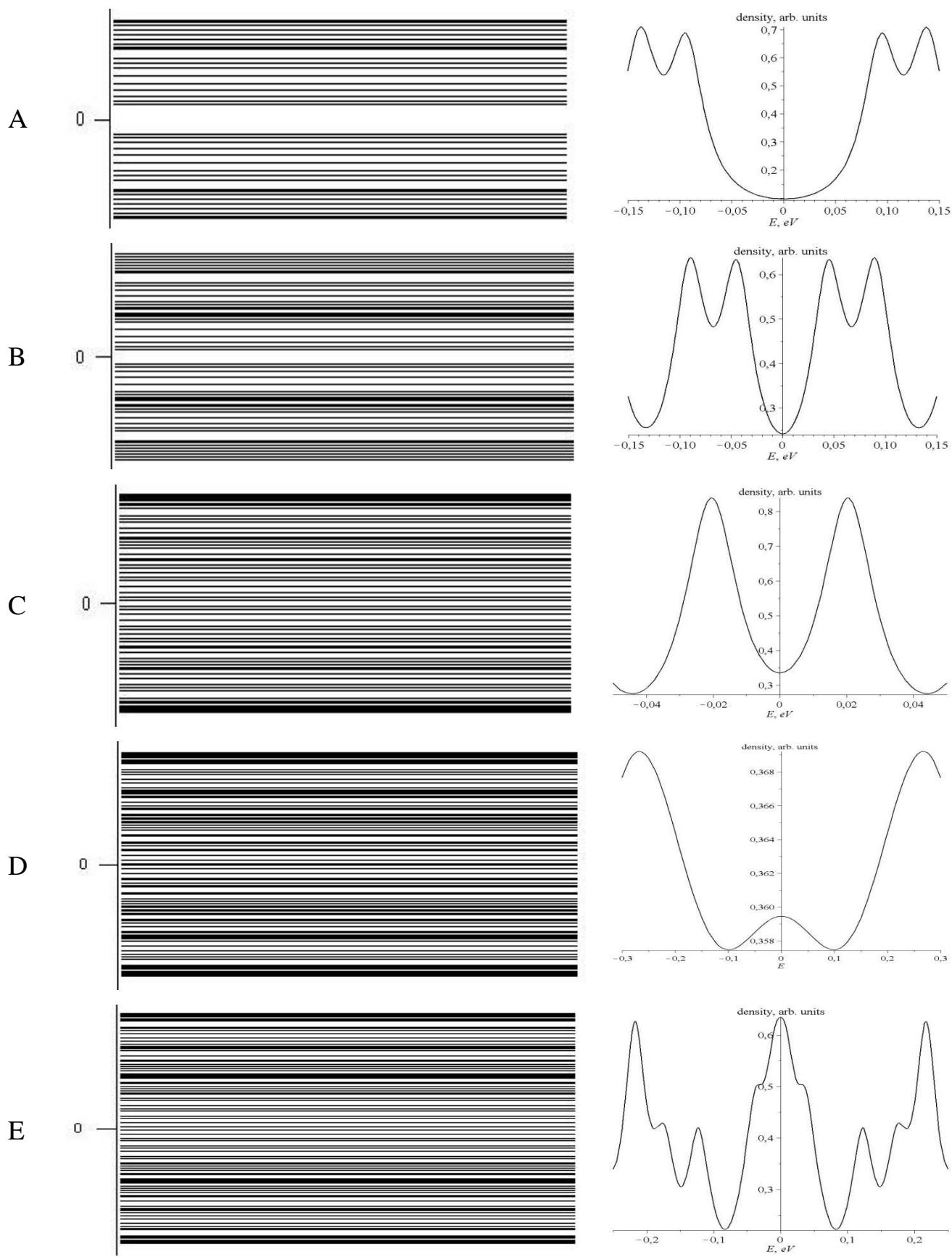

**Figure 6.** Energy spectrum near the Fermi level (left), electron state density (right), A – 9-BPR, B – 12-BPR, C – 15-BPR, D – 18-BPR, E – 21-BPR

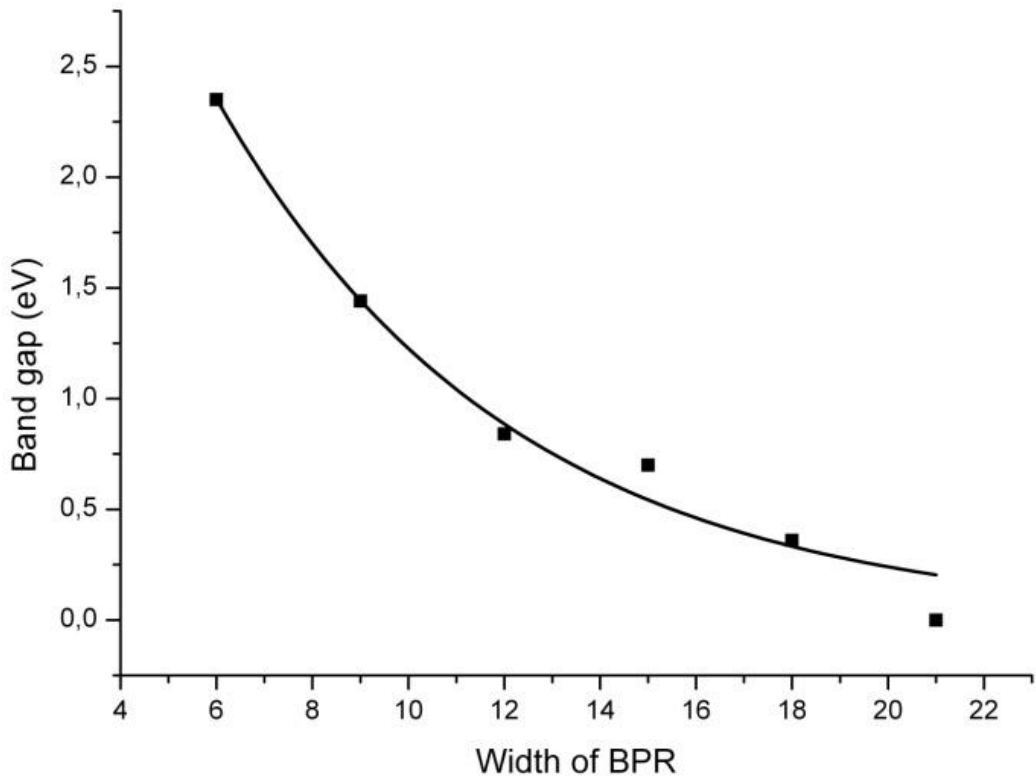

A

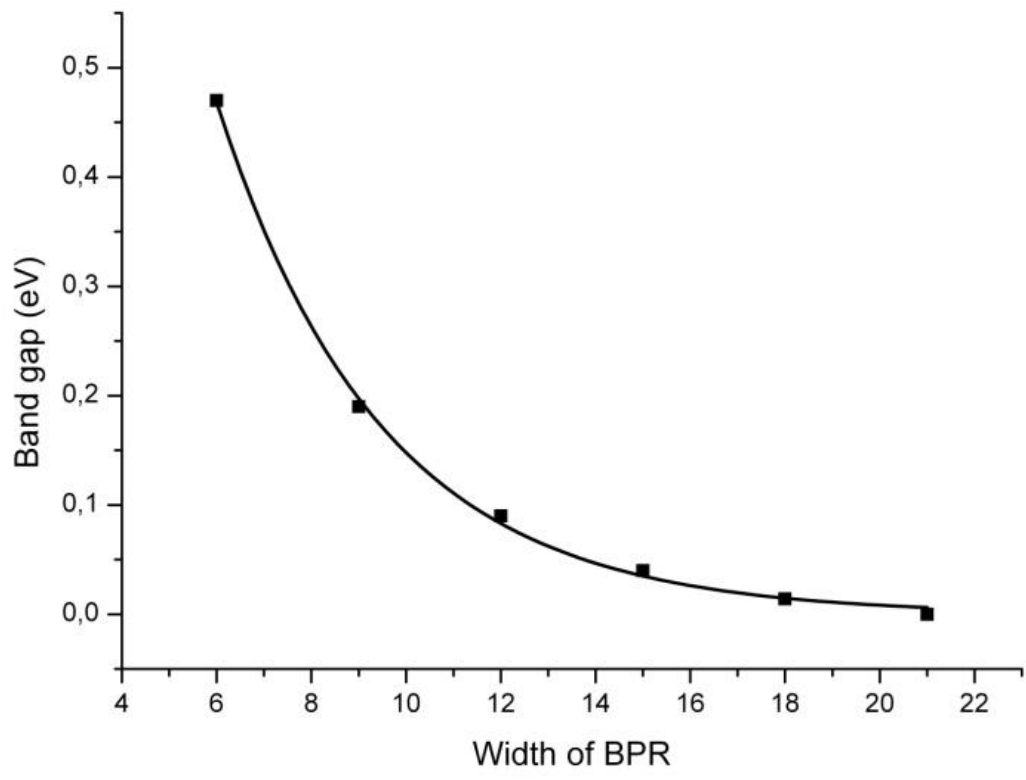

B

**Figure 7.** Dependence of the band gap width on the width of the biphenylene ribbon, A – according to the results of [5], B – according to theoretical results